\documentclass[fleqn,usenatbib]{mnras}

\usepackage{newtxtext,newtxmath}

\usepackage[T1]{fontenc}

\DeclareRobustCommand{\VAN}[3]{#2}
\let\VANthebibliography\thebibliography
\def\thebibliography{\DeclareRobustCommand{\VAN}[3]{##3}\VANthebibliography}

\usepackage{graphicx}	
\usepackage{amsmath}	
\usepackage{rotating}

\newcommand{\msol}{\mathcal{M}_\odot}

\newcommand{\sfe}{\mathrm{SFE_{gl}}}

\title[Evolution of clusters with or without black holes]{Evolution of open clusters with or without black holes}

\author[B. Shukirgaliyev et al.]{B.~Shukirgaliyev$^{1,2}$\thanks{E-mail: bekdaulet.shukirgaliyev@nu.edu.kz (BS)},
P.~Berczik$^{3,4,5}$,
A.~Otebay$^{6,2}$,
M.~Kalambay$^{7,2,1}$,
A.~Kamlah$^{8,3}$,
Y.~Tleukhanov$^{6,1}$,
\newauthor
E.~Abdikamalov$^{6,1}$,
S.~Banerjee$^{9,10}$,
A.~Just$^{3}$\\
$^{1}$Energetic Cosmos Laboratory, Nazarbayev University, 53 Kabanbay Batyr ave., 010000 Astana, Kazakhstan\\
$^{2}$Fesenkov Astrophysical Institute, 23 Observatory str., 050023 Almaty, Kazakhstan\\
$^{3}$Astronomisches Rechen-Institut, Zentrum f\"ur Astronomie der Universit\"at Heidelberg, M\"onchhofstr 12-14, 69140 Heidelberg, Germany\\
$^{4}$Konkoly Observatory, Research Centre for Astronomy and 
Earth Sciences, E\"otv\"os Lor\'and Research Network (ELKH), \\
MTA Centre of Excellence, Konkoly Thege Mikl\'os \'ut 15-17, 
1121 Budapest, Hungary\\
$^{5}$Main Astronomical Observatory, National Academy of Sciences of Ukraine, 27 Akademika Zabolotnoho St., 03143 Kyiv, Ukraine\\
$^{6}$Physics Department, Nazarbayev University, 53 Kabanbay Batyr ave., 010000 Astana, Kazakhstan\\
$^{7}$Al-Farabi Kazakh National University, 71 Al-Farabi ave., 050040 Almaty\\
$^{8}$Max-Planck-Institut f\"ur Astronomie, K\"onigstuhl 17, 69117 Heidelberg, Germany\\
$^{9}$Helmholtz-Institut f\"ur Strahlen- und Kernphysik, Nussallee 14-16, D-53115 Bonn, Germany\\
$^{10}$Argelander-Institut f\"ur Astronomie, Auf dem H\"ugel 71, D-53121 Bonn, Germany\\
}

\date{Accepted XXX. Received YYY; in original form ZZZ}

\pubyear{2022}

\begin{document}
\label{firstpage}
\pagerange{\pageref{firstpage}--\pageref{lastpage}}
\maketitle

\begin{abstract}
Binary black holes (BHs) can be formed dynamically in the centers of star clusters. The high natal kicks for stellar-mass BHs used in previous works made it hard to retain BHs in star clusters. Recent studies of massive star evolution and supernovae (SN) propose kick velocities that are lower due to the fallback of the SN ejecta. We study the impact of these updates by performing $N$-body simulations following instantaneous gas expulsion. For comparison, we simulate two additional model sets with the previous treatment of stars: one with high kicks and another with artificial removal of the kicks. Our model clusters initially consist of about one hundred thousand stars, formed with centrally-peaked efficiency. We find that the updated treatment of stars, due to the fallback-scaled lower natal kicks, allows clusters to retain SN remnants after violent relaxation. The mass contribution of the retained remnants does not exceed a few percent of the total bound cluster mass during the early evolution. For this reason, the first giga year of evolution is not affected significantly by this effect. Nevertheless, during the subsequent long-term evolution, the retained BHs accelerate mass segregation, leading to the faster dissolution of the clusters. 
\end{abstract}




\section{Introduction}

Recent observations of the gravitational wave (GW) signal from the binary black hole (BH) merger GW190521 by the LIGO and Virgo observatories filled the upper-mass gap forbidden for stellar-mass BHs with a final merger mass of $142^{+28}_{-16}M_\odot$ \citep{BBH150-2020PhRvL, BBH150-2020ApJL}. Such high mass BHs may form via hierarchical mergers of BHs in dense stellar systems \citep{AntoniniRasio16, GerosaBerti17, Stone+17, McKernan+18, Rodriguez+19, Yang+19, ArcaSedda+20, Mapelli+22, Torniamenti+22}. Due to dynamical friction, the BHs in stellar clusters tend to sink to the central part \citep{1993Natur.364..421K,1993Natur.364..423S}. This results in mass segregation, which may lead to the dynamical formation of binaries and mergers \citep{dragon2016,Askar+2018, Kremer+2020ApJS..247...48K}. However, due to asymmetric core-collapse supernova (ccSN) explosion, supernova (SN) remnants may form with natal kicks \citep{Fryer2004, bray16, janka17, mandel20, Kapil23}. If the BHs receive sufficiently strong kicks, they escape the star cluster before the mass segregation brings them to the cluster center \citep{Ernst+2011,Kremer+2018}. At the same time, stellar-mass BHs have been observed in extragalactic \citep{2007Natur.445..183M,2011ApJ...734...79B} and Galactic \citep{2012Natur.490...71S,2013ApJ...777...69C, 2015MNRAS.453.3918M,2018ApJ...855...55S} globular clusters as well as in young massive clusters of the Large Magellanic Cloud \citep{2022MNRAS.511.2914S,2022NatAs...6.1085S,2022A&A...665A.180L}.

There have been multiple simulations of the evolution of stellar clusters with different prescriptions for stellar evolution and supernova explosion. These prescriptions affect a wide range of stellar parameters, including the natal kick velocities of the supernova remnants. In simulations that assume as high a natal kick for BHs as those for neutron stars  \citep[NS,][]{JonkerNelemans2004, Hobbs+2005, Repetto+2012}, stellar clusters struggle to retain the BHs \citep{Ernst+2011,Kremer+2018}. To retain the observed fraction of BHs in globular clusters \citep{SollimaBaumgardt2017, BaumgardtSollima2017}, different groups resorted to the artificial reduction of the BH natal kick velocities \citep{Peuten+2016, Pavlik+2018, Kremer+2018,2020MNRAS.492.2936A}. However, recent advances in stellar evolution and supernova explosions suggest that such artificial reductions may not be necessary to obtain lower natal kicks \citep{2017NewAR..78....1M, Kremer+2020ApJS..247...48K}. 
 
Some core-collapse supernovae experience fallback of the ejecta material back to the supernova remnant \citep{janka22}. This reduces the natal kick velocity of the latter \citep{janka17}. Detailed modeling of stellar evolution and supernova explosions are long-standing problems in astrophysics \citep[][for a recent review]{mueller20review}. The prescriptions used in $N$-body codes are based on approximate models based on simulations and/or observational constraints. In particular, prescriptions of stellar evolution and SN used in earlier $N$-body simulations \citep[e.g.][]{Ernst+2011} are based on the approximate model of \citet{Polsetal1998} and do not include this fallback-scaling of the kick velocities. In the literature, these prescriptions are called as SSE for single stars \citep{Hurley+00,sse13} and BSE for stars in close binaries \citep{Hurley+02,bse13}. 

Recently, \citet{Kamlah+2021} updated the SSE and BSE prescriptions for the NBODY6++GPU code \citep{wang+2015}. The updates include metallicity-dependent stellar winds \citep{Vink_2001, Belczynski+2010}, ccSNe \citep{Fryer+2012, Banerjee+2020}, fallback-scaled natal kicks \citep{MeakinArnett2006, MeakinArnett2007, FryerYoung2007, Scheck2008, Fryer+2012, Banerjee+2020}, updated treatment of electron-capture SNe, accretion- and merger-induced collapse remnant masses and natal kick \citep{Kiel+2008, GessnerJanka2018}, pair-instability SNe \citep{Belczynski+2016, Woosley2017}, and BH natal spins \cite{belczynski20, belczynski2020b}. Similar efforts were made in the context of other $N$-body codes \citep[see ][]{mobse18, Banerjee+2020, Banerjee2021}. Recently, \citet{Bek+21} implemented these SSE prescriptions in the direct $N$-body code phi-GRAPE/GPU \citep{Harfst+07, Ernst+2011, Just2012, Berczik+13}. In particular, these changes allow us to take into account the fallback-scaled natal kicks for stellar-mass BHs. This creates a channel for increasing the BH retention fraction of star clusters without artificial reduction of the kick velocities.  

In this work, we study the impact of the different SN natal kick treatments on the evolution of star clusters. We assume that our model clusters are formed with a centrally-peaked star-formation efficiency \citep[SFE,][]{Bek+17}. We perform three sets of simulations with the same initial conditions. In the first set, we use the updated SSE \citep{Kamlah+2021, Bek+21} that results in a lowered natal kick for some fraction of BHs and NSs due to the fallback of the explosion material. In the second set, we use  the old SSE that results in high natal kicks \citep{Ernst+2011}. Finally, in the third set, we again use the old SSE but artificially neglect the kicks \cite[as in ][]{Bek+17, Bek+18, Bek+19}. We study how these changes in stellar evolution and supernova prescriptions alter the outcome of the cluster evolution after instantaneous gas expulsion (IGE). 

This paper is structured as follows. We describe our star cluster model and $N$-body simulations in sec.~\ref{sec:methods}. We present our new results on the effect of natal kick on the evolution of model clusters in sec.~\ref{sec:results}. Section~\ref{sec:con} presents our conclusions and discussions.

\section{Model and Simulations}
\label{sec:methods}

Following \citet{Bek+17}, we use star cluster models with a \citet{Plummer_1911} density profile prior to IGE. We assume that our model star clusters are formed with a constant efficiency per free-fall time in a centrally-concentrated spherically symmetric clump gas according to the semi-analytic model of \citet{PP13}. In such clusters, stars are more concentrated in the cluster center than the residual gas, having a steeper density profile. Consequently, the star cluster can survive with a global SFE as low as 15 percent \citep{Bek+17}, almost independent of the impact of the Galactic tidal field \citep{Bek+19}. 

We consider star clusters consisting of $N_\star = 104554$ stars with an initial stellar mass of $M_\star=6\cdot10^4\msol$ at the time of IGE. The model clusters were formed with different global SFEs as 0.15, 0.17, 0.20, and 0.25. By the global SFE, we mean the gas mass fraction within ten Plummer scale radii, $10a_\star,$ converted into stars before IGE. We generate the initial conditions assuming virial equilibrium within the gravitational potential of stars and the residual gas. The density profiles of the gas for a given global SFE are calculated according to the \citet{PP13} model. Our $N$-body simulations start at the time of IGE when star clusters are in a supervirial dynamical state. We use the stellar initial mass function (IMF) of  \citet{Kroupa2001} within the mass limit of $0.08 \msol \leq m_\star \leq 100 \msol$. We assume that all stars in the simulation have solar metallicity $(Z=0.02)$. More details of the initial conditions can be found in \citet{Bek+17}.

Our model clusters orbit in the Galactic potential given by a three-component axisymmetric Plummer-Kuzmin model \citep{MiyamotoNagai75,Just+09} with the same parameters as in \citet{Bek+21}. The model clusters are moving on a circular orbit at the solar galactocentric distance of $R_\mathrm{orb}=8\,178$~pc \citep{GravityColl2019} in the disk plane {with the orbital speed of $V_\mathrm{orb}=234.73\ \mathrm{km\ s^{-1}}$}. We calculate the Jacobi radius using Eq.~13 from \citet{Just+09}:
\begin{equation}\label{eq:rj}
    r_\mathrm{J}=\left(\displaystyle\frac{GM_\star}{\left(4-\beta^2\right)\Omega^2}\right)^{1/3},
\end{equation}
where $G$ is the gravitational constant, $\beta=1.37$ is the normalized epicyclic frequency, $\Omega=V_\mathrm{orb}/R_\mathrm{orb}$ is the angular speed of the star cluster on a circular orbit. We obtain an initial value of $r_\mathrm{J}=52.84\ \mathrm{pc}$ for the above parameters. The half-mass radius of the cluster, which encloses half of the bound stellar mass of the cluster, is assumed to be $5\%$ of the Jacobi radius, which results in $r_\mathrm{h}=2.64\ \mathrm{pc}$. 

We assume that stars are gravitationally bound to the cluster if they reside within the Jacobi radius, $r_\mathrm{J}$, regardless of their kinetic energies. We define the gravitationally bound mass (also we call it Jacobi mass, $M_\mathrm{J}$) and the Jacobi radius iteratively for each snapshot using Eq.~(\ref{eq:rj}), where we substitute $M_\star$ with
\begin{equation}\label{eq:mj}
    M_\mathrm{J} = \sum{m_\star(<r_\mathrm{J})},
\end{equation}
with  current masses, $m_\star$, of individual stars in the cluster.

We perform simulations using the direct $N$-body code phi-GRAPE/GPU \citep{Harfst+07, Ernst+2011, Just2012}. We run until 12.5 Gyr, which represents a typical age of globular clusters \citep{VandenBerg+2013}, or until the cluster mass decreases below $50\ \msol$. In our simulations, the Galactic center is the origin of the reference frame. We keep all stellar objects in the simulations, even if they escaped the cluster. When the mean density of the  cluster becomes comparable to that of the Galactic field, the center of the cluster is not well defined any longer. Therefore we stop our simulations when the Jacobi mass of a cluster drops below $50\msol$.

We do not follow binaries in our simulations. Instead, we use a gravitational potential softening with a scaling radius of $\epsilon\approx 41\ \mathrm{AU}$.
However, this did not stop the formation of multiple systems of BHs at the cluster center in some simulations. Therefore some simulations did not finish yet. The list of objects in such artificial compact systems will be provided at the end of Sec.~\ref{sec:long-term}. 

In the following, the first set of simulations, which uses the updated SSE, is referred to as `{\it new SSE}.' The second set, which uses the old SSE and has a high natal kick drawn from a Maxwellian distribution with velocity dispersion $265\ \mathrm{km\ s^{-1}}$ \citep{Hobbs+2005}, is referred to as `{\it old SSE}.' Finally, the third set, which uses the old SSE with no kicks, is referred to as {\it old SSE, no kick}.'

\section{Results}
\label{sec:results}

\begin{figure}
    \includegraphics[width=\linewidth]{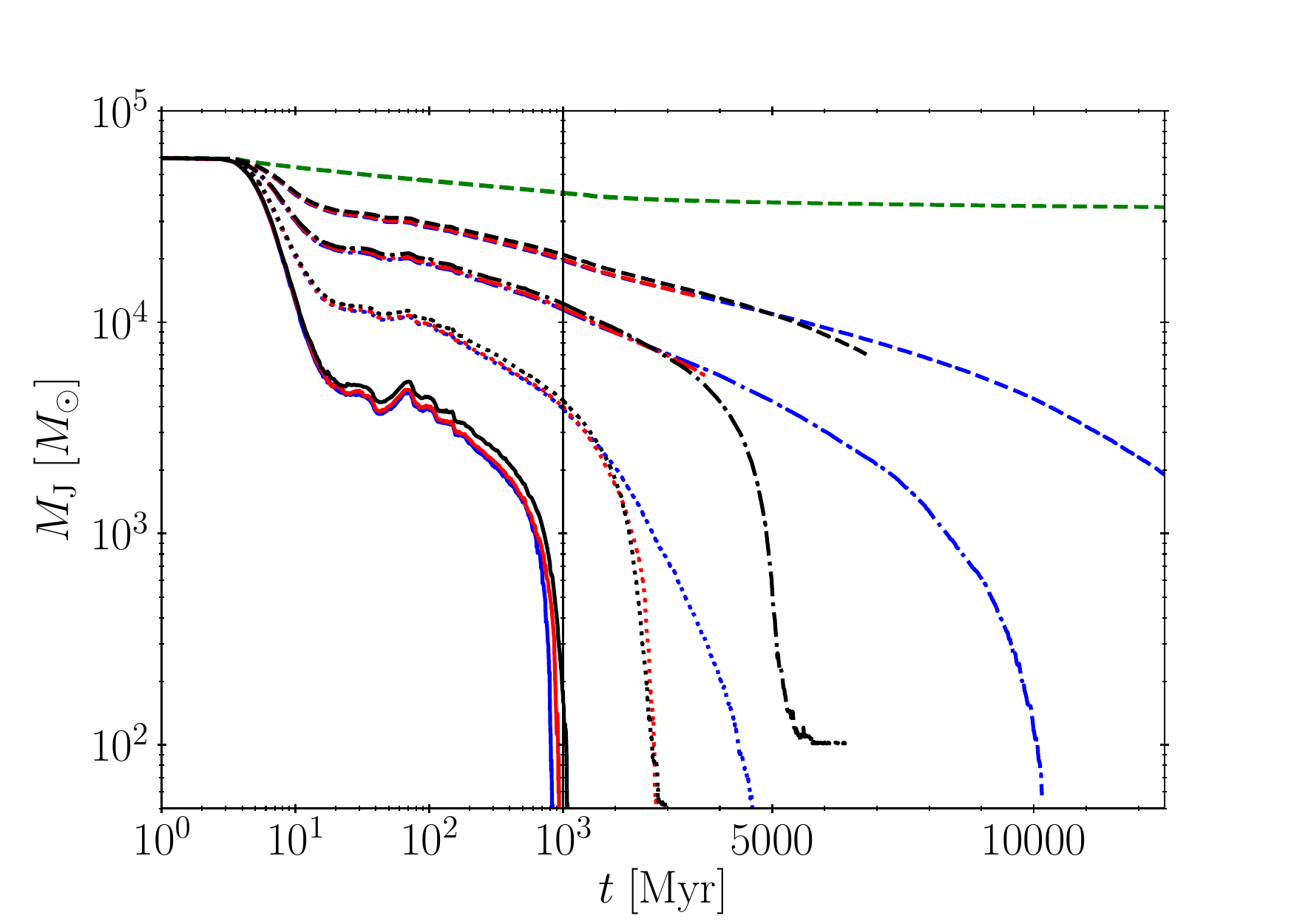}
    \includegraphics[width=\linewidth]{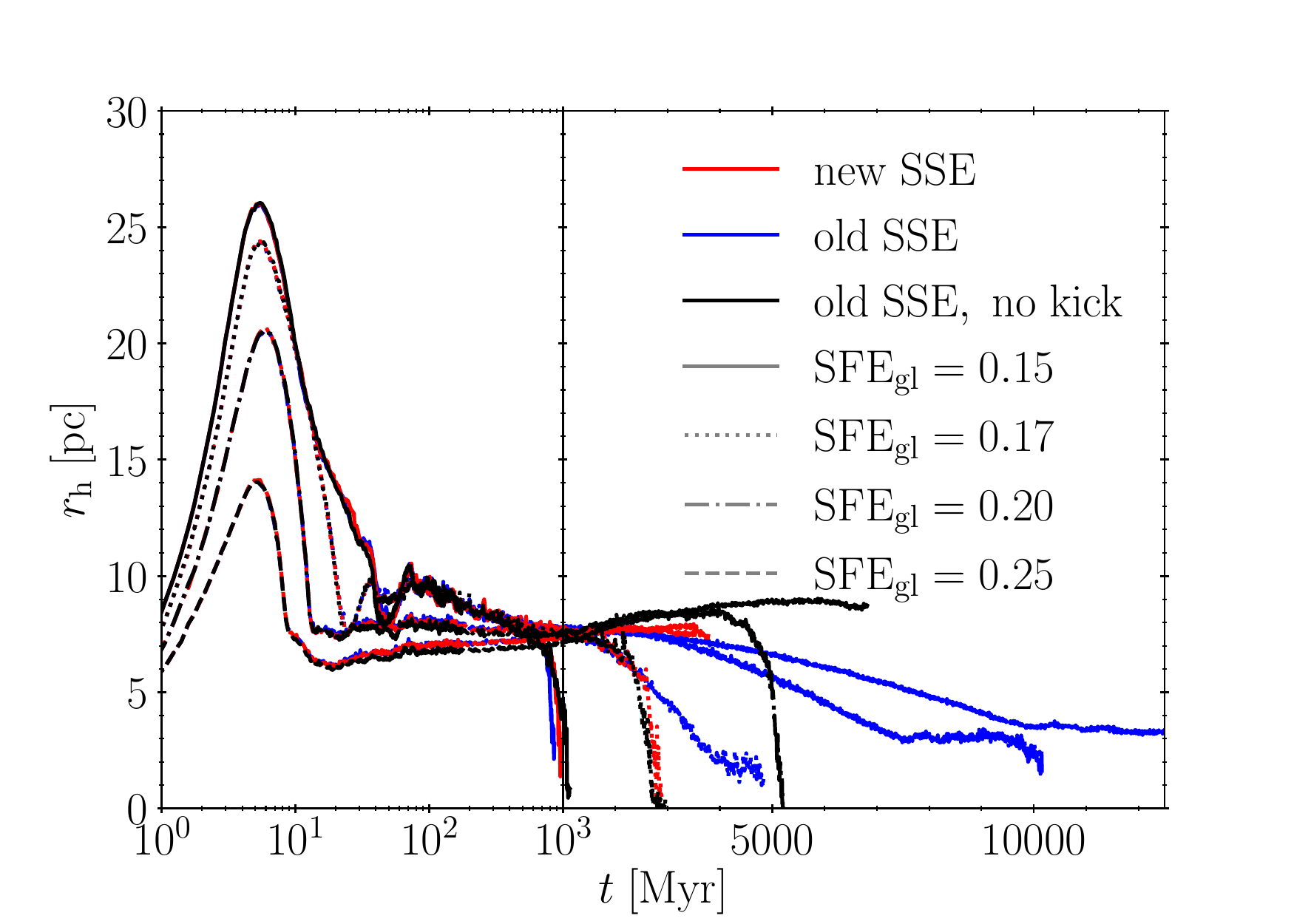}
    \includegraphics[width=\linewidth]{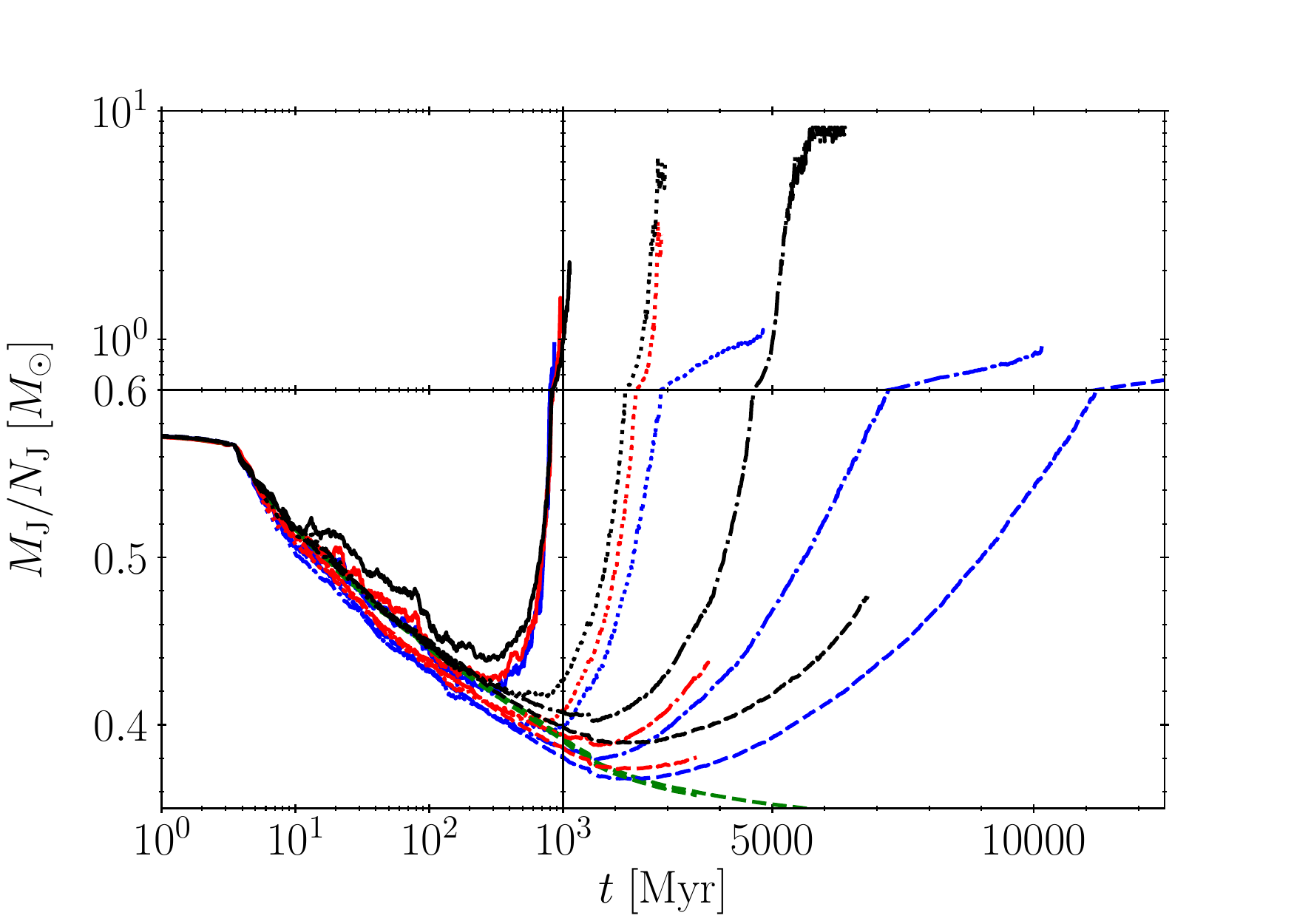}
    \caption{Jacobi mass $M_\mathrm{J}$ (top panel), half-mass radius $r_\mathrm{h}$ (middle panel, and the mean stellar mass $M_\mathrm{J}/N_\mathrm{J}$ (bottom panel) as a function of time for different cluster models. The red, blue, and black lines represent the three sets of simulations. The green line represents the total stellar mass, including both bound and escaped stars. Different line styles indicate global SFEs of 0.15, 0.17, 0.20, and 0.25. The $x$-axis is in log (linear) scale below (above) $10^3$~Myr. The $y$-axis of the bottom panel has a linear (log) scale below (above) $0.6M_\odot$.}
    \label{fig:mass}
\end{figure}

Since our $N$-body simulations start after IGE, our model clusters expand immediately in the beginning \citep{BK07,PB2012,Bek+17,Brinkmann+17}. Figure~\ref{fig:mass} shows the Jacobi mass (top panel), half-mass radius (middle panel), and the mean stellar mass (bottom panel) as a function of time. The mean stellar mass\footnote{In our calculation of mean stellar mass, we consider the contributions of not only stars but also stellar remnants, including white dwarfs, neutron stars, and black holes. This applies to our calculation of the total mass as well.} of a cluster is defined as the ratio of bound mass $M_\mathrm{J}$ to the number of bound stars $N_\mathrm{J}$ (i.e., within the Jacobi radius). The red, blue, and black lines represent the three simulation sets we outline in the figure key. The line styles correspond to different SFEs. The green dashed line in the top panel represents the total stellar mass, including all bound and escaped stellar objects. To see the evolutionary tracks of star clusters in both small and large time scales, we use a logarithmic (linear) scale for the $x$-axis before (after) 1 Gyr timescale, shown by the vertical line. The linear scale after 1 Gyr allows to see the difference in the cluster dissolution dynamics. The vertical axis of the bottom panel is divided into two parts with linear (logarithmic) scales below (above) $0.6 M_\odot$.

The cluster half-mass radius expands considerably in the first few Myr of evolution. In this period, the cluster mass loss happens mainly due to stellar winds, which manifests as a shallow plateau in the average stellar mass seen in the bottom panel of Fig.~\ref{fig:mass}. The supernova explosions set in after $\sim 3$ Myr, accelerating the decrease of average stellar mass. The bound mass rapidly decreases after $\sim 5$ Myr, when the first escaping stars cross the Jacobi radius. This caused a rapid decrease of the half-mass radius until the cluster reached its new quasi-equilibrium state at $\gtrsim 20$ Myr. We can see this as a plateau in the time evolution of the bound mass. In light of this result, in the following, we assume that the violent relaxation -- defined as a dynamical response to gas expulsion -- finishes at $t=20\ \mathrm{Myr}$, as in \citep{Bek+17}. The overall dynamics of the cluster during the violent relaxation that we observe in our simulations are consistent with previous studies \citep{GeyerBurkert2001,BK07,Smith+11,Bek+17,BekPhDT18,Brinkmann+17}.

The sub-group of unbound stars with the lowest kinetic energy may reside in the near-circular orbits around the parent cluster \citep{Ross+1997MNRAS.284..811R,FH2005,Ernst+2015}. While these are energetically unbound (the Galactic potential drives their dynamics), they may periodically cross the Jacobi radius due to their epicyclic motion \citep{Kuepper+2008,Just+09}. When this happens, their masses are added to our bound mass calculations. We see this as minor oscillations in our model cluster bound masses and half-mass radii occurring after 20 Myr. Due to their lower stellar density, these oscillations are more pronounced in clusters with the lowest global SFE.  

Once the cluster reaches its new virial equilibrium, its evolution is mostly driven by two-body relaxation \citep{1999PhDT.........1E,Baumgardt2001,bm2003,TF2005,GB2008} and Galactic tidal stripping \citep{Whitehead+2013,Ernst+2015} mechanisms \citep{Bek+18}. During this phase, stars exchange their energies via close encounters. High-mass stars transfer their kinetic energies to low-mass stars. The low-mass stars are kicked out of the cluster, while high-mass objects sink into the cluster center, resulting in mass segregation \citep{1993Natur.364..421K,1993Natur.364..423S}. This is observed as an increase in the mean stellar mass within star clusters, as seen in the bottom panel of Fig.~\ref{fig:mass}.

In the following, we will refer to the dynamics before (after) 1 Gyr as early  (long-term) evolution, and we will discuss the impact of SSE updates separately for these two phases. 

\subsection{Early evolution}

\begin{figure}
    \includegraphics[width=\linewidth]{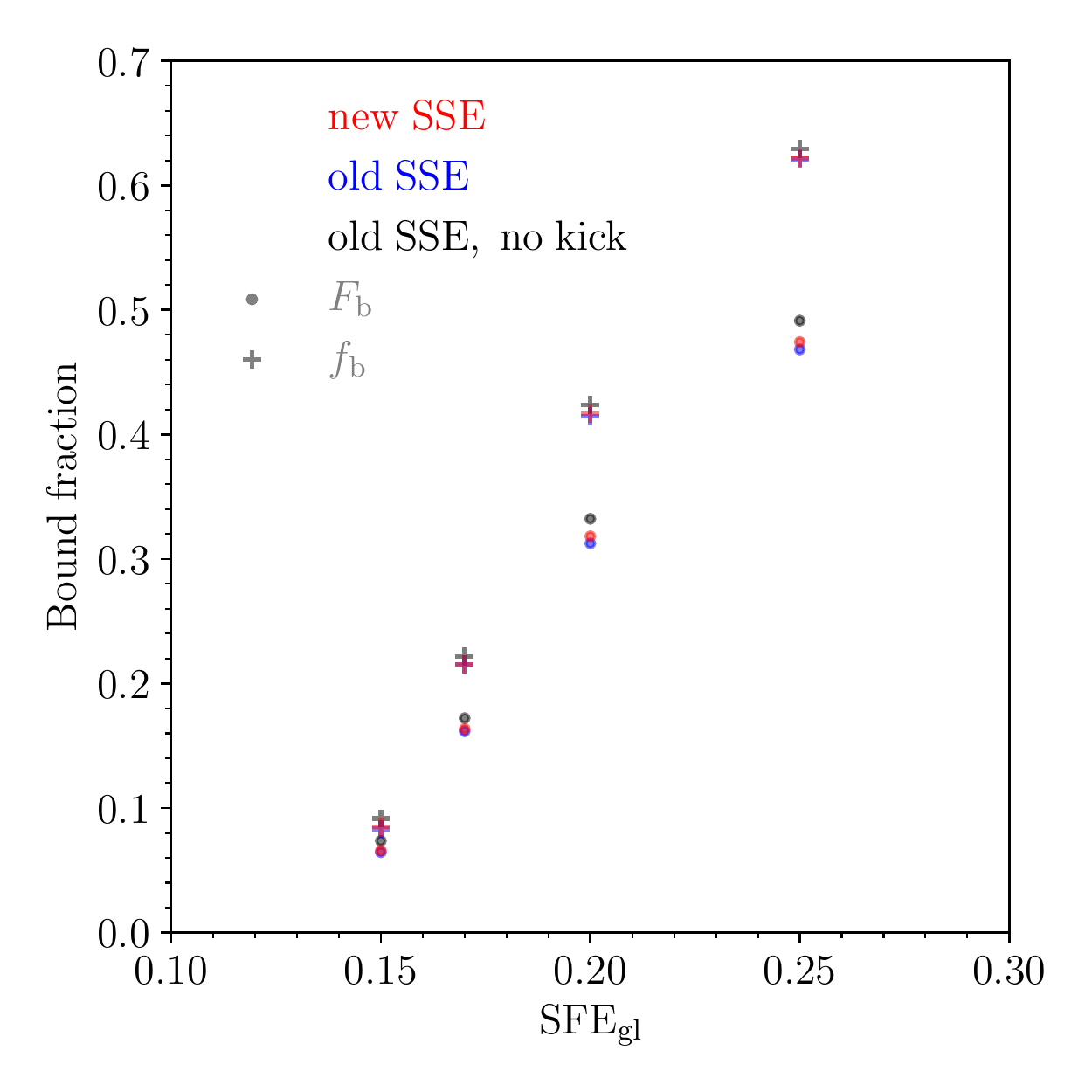}
    \caption{Mass (dots) and number (pluses) fractions of bound objects at $100\ \mathrm{Myr}$ after IGE as a function of global SFE.
    }
    \label{fig:Fb}
\end{figure}

\begin{figure}
	\includegraphics[width=\linewidth]{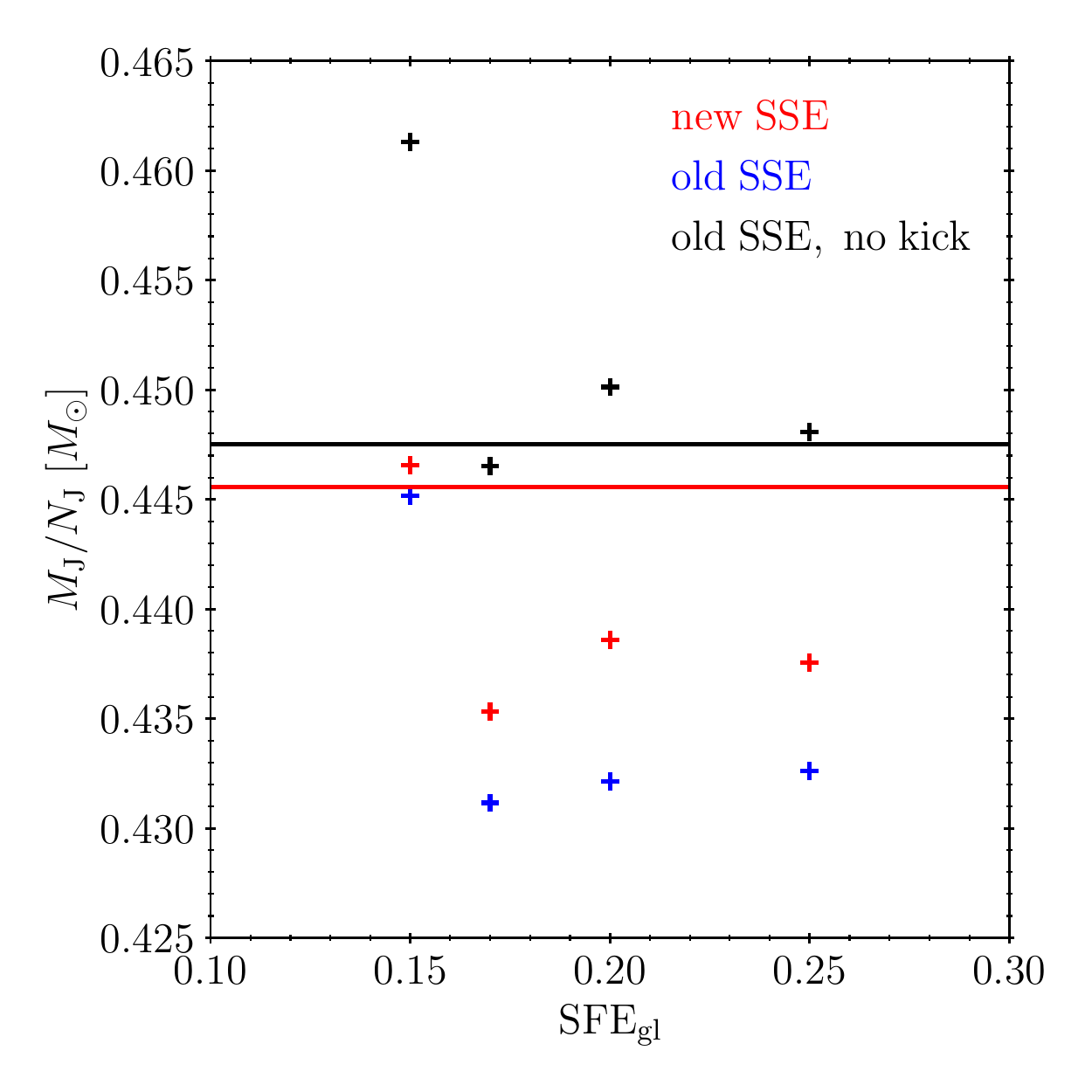}
    \caption{Mean stellar mass of model clusters at $100\ \mathrm{Myr}$ after IGE as a function of global SFE. The horizontal lines show the mean mass of all stellar objects, including both bound and unbound stars and remnants for the old and new SSE.}
    \label{fig:mm100}
\end{figure}

\begin{figure}
	\includegraphics[width=\linewidth]{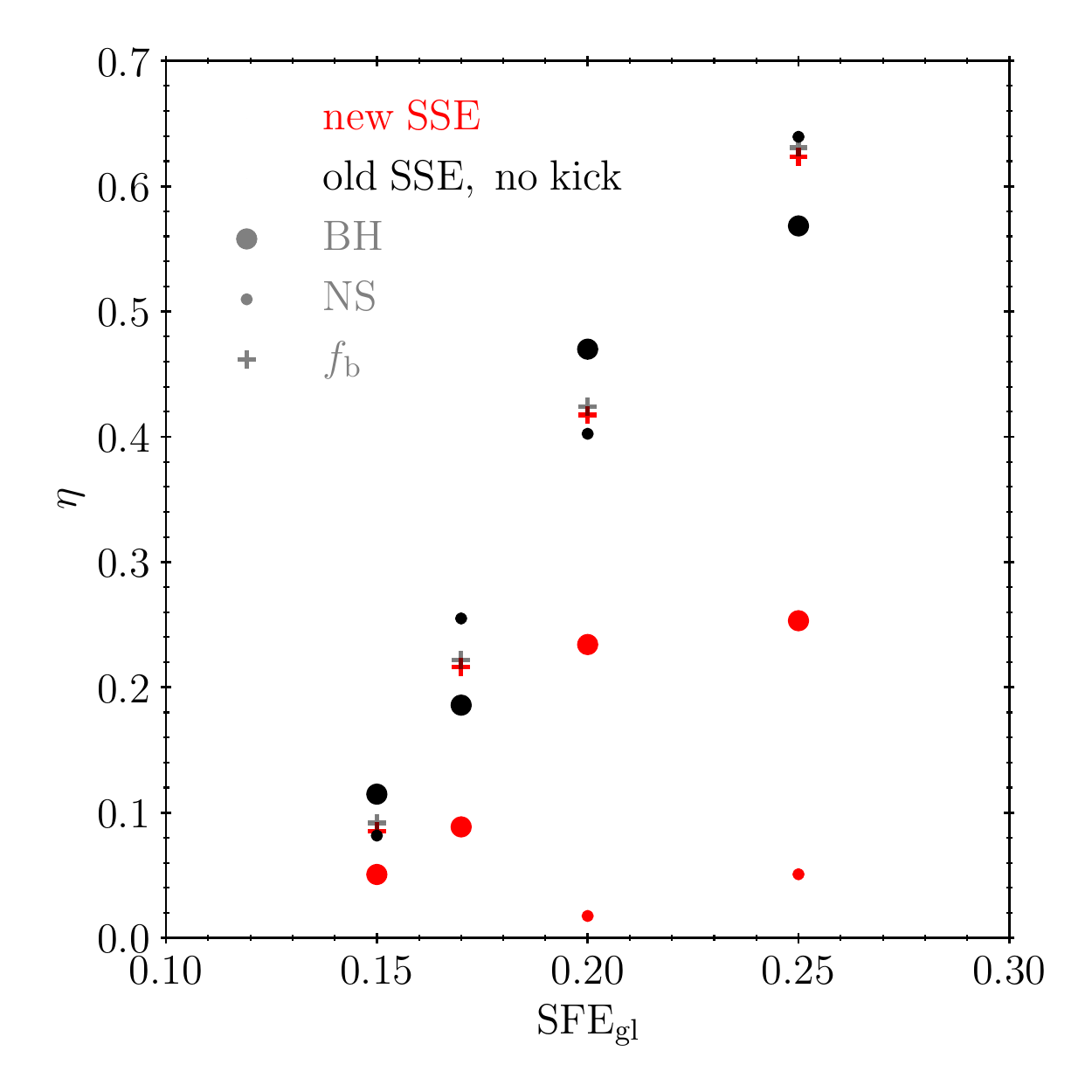}
    \caption{BH and NS retention fractions for models with old SSE neglecting natal kicks and new SSE with fallback-scaled natal kicks. Colors correspond to the model set, and symbols show the retention fractions of BH, NS, and all stars at $t=100$~Myr, as indicated in the key. 
    }
    \label{fig:eta}
\end{figure}

Fig.~\ref{fig:mass} shows that the evolutionary tracks of models from separate sets do not differ much from each other in the early phase. The most significant differences among models with the same $\sfe$ are observed in the mean stellar masses, as can be seen in the bottom panel of Fig.~\ref{fig:mass}. 

We compare the cluster properties in more detail at $t=100\ \mathrm{Myr}$, which is well after the violent relaxation but early enough to be separated from the long-term evolution. All our cluster models are in a quasi-equilibrium state at this time frame. Fig.~\ref{fig:Fb} shows the fractions of bound mass $F_\mathrm{b}$ and number $f_\mathrm{b}$ as a function of $\sfe$. The bound mass (number) fraction is the ratio of the bound mass (number) of objects to the total stellar mass (number) immediately after the IGE:
\begin{eqnarray}
F_\mathrm{b}=\frac{M_\mathrm{J}(t=100\ \mathrm{Myr})}{M_\star},\\
f_\mathrm{b}=\frac{N_\mathrm{J}(t=100\ \mathrm{Myr})}{N_\star}.
\end{eqnarray}
As we can see, these quantities are within a few percent of each other for any given value of the $\sfe$. 

Figure~\ref{fig:mm100} shows the mean stellar mass as a function of $\sfe$. The old treatment of SSE with high (no) natal kicks yields the lowest (highest) mean masses, while the new SSE yields values that are in between. The mean stellar mass reflects the retention fraction of SN remnants. In particular, models neglecting SN natal kicks have the highest mean stellar masses (black), indicating the highest retention of SNe remnants. On the other hand, models with high natal kicks eject all SN remnants, resulting in low mean stellar masses (blue). For all values of $\sfe$ except that of 0.15, the mean masses in clusters without kicks are similar to the mean mass of all stellar objects, including escaped stars, shown by horizontal solid lines. The case of $\sfe=0.15$ represents the minimum value for cluster survival in the aftermath of IGE. Due to their low stellar density at the end of violent relaxation, the highest-mass objects are more likely to remain in the cluster \citep{Bek+18}, which yields higher mean stellar masses. 

The differences in the bound mass (number) fractions seen in Fig.~\ref{fig:Fb} can be explained by the retention fraction $\eta$ for NSs and BHs, shown in Fig.~\ref{fig:eta}. The retention fraction $\eta$ is the fraction of all formed BHs or NSs remaining within the Jacobi radius. Since no SN remnants remain in models with high natal kicks, we omit this set in the comparisons. The retention fractions of NSs and BHs follow the same trend against $\sfe$ as the bound number fraction in case of "old SSE, no kicks". The retention fraction of NSs is below $5\%$ for the new SSE models. The retention fraction of BHs is consistent with the number of BHs that receive low fallback-scaled kicks. 

In summary, the early evolution of cluster with new and old SSE is similar to each other because the total mass contribution of NSs and BHs is of the order of a few percent of the total cluster mass. However, the situation changes when we look at the long-term evolution.

\subsection{Long-term evolution}\label{sec:long-term}

\begin{figure}
	\includegraphics[width=\linewidth]{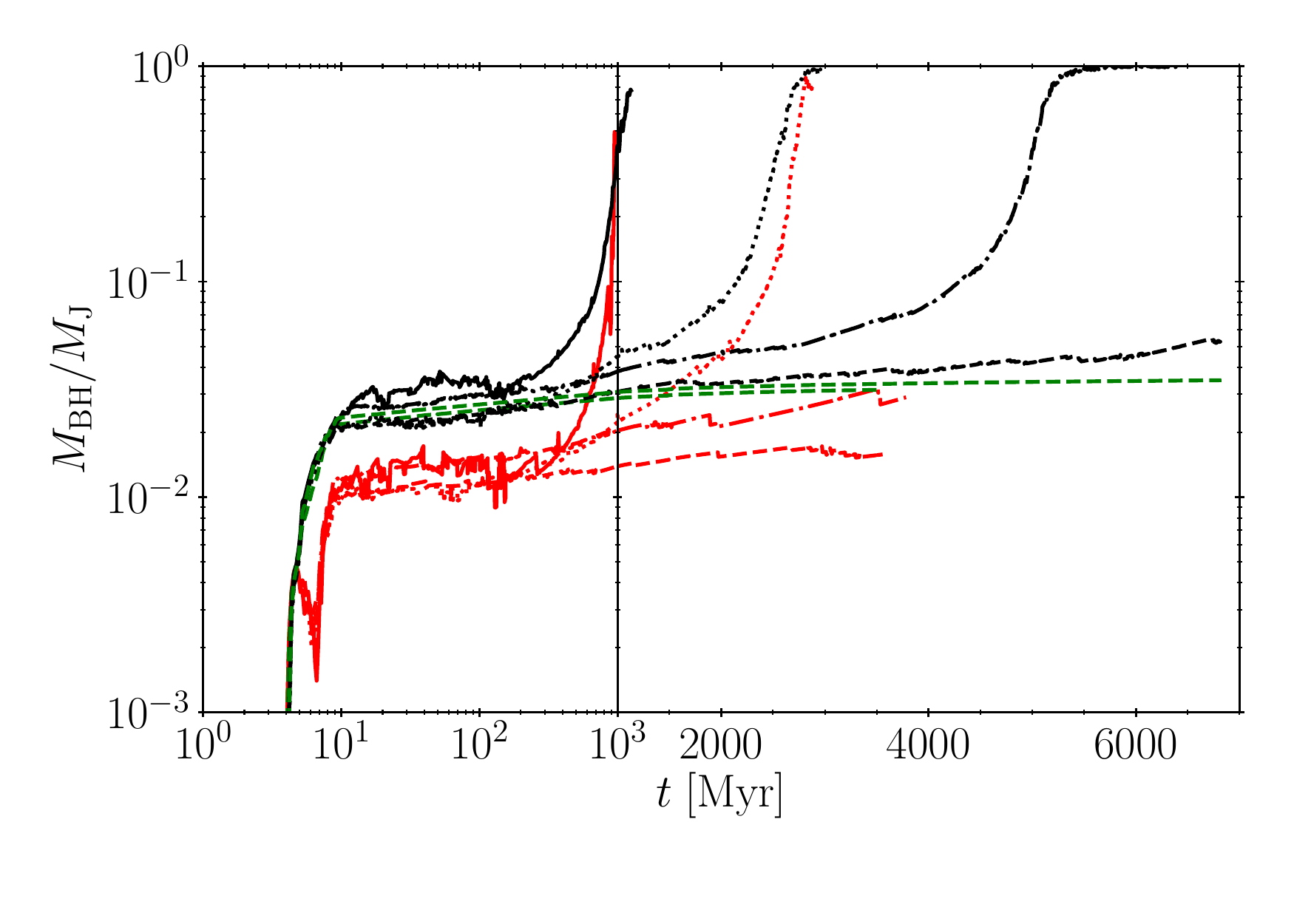}\\
        \includegraphics[width=\linewidth]{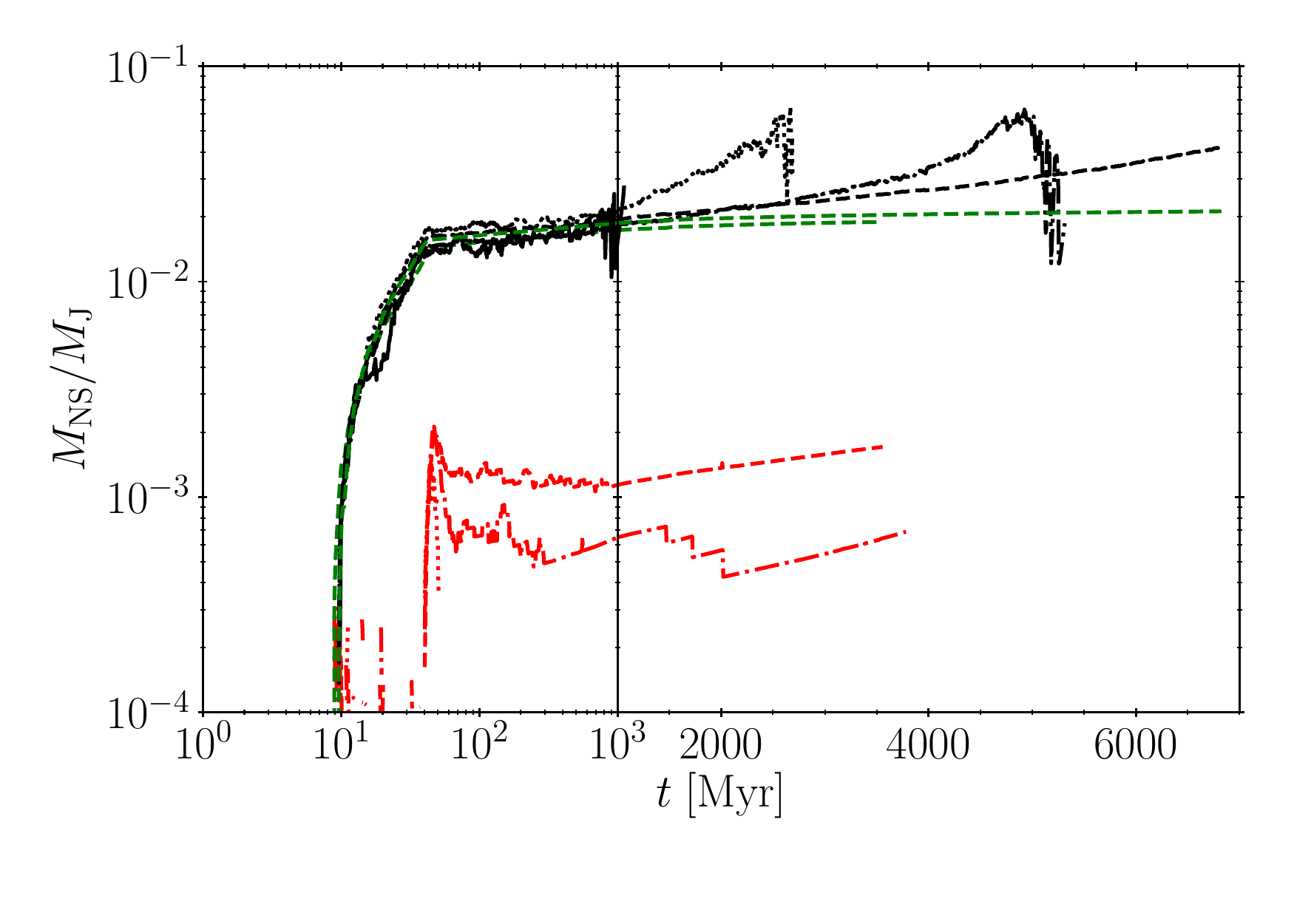}
    \caption{Fractions of bound masses of BHs (top panel) and NSs (bottom panel) as a function of time. Line styles and color-coding as in Fig.~\ref{fig:mass}. We omit the `old SSE' models because they lack SN remnants due to high natal kicks. The green lines show the fraction of mass in BHs and NSs if the model clusters would evolve, keeping all stars in.}
    \label{fig:longtermBHs}
\end{figure}

Although the `new SSE' models behave closer to `old SSE' models during the early evolution, their long-term dynamics are closer to the `old SSE, no kick' models. That is because the presence of BHs as massive objects significantly influences the long-term evolution of star clusters by accelerating mass segregation. This, in turn, accelerates the dissolution of star clusters.  
It is illustrated in the mean stellar mass of the model clusters (bottom panel of Fig.~\ref{fig:mass}), where black and red lines corresponding to `old SSE, no kicks' and `new SSE' model sets can go very high within a short timescale. In contrast, the blue lines never reach $1M_\odot$ because these clusters do not contain any NS or BH. 

Figure~\ref{fig:longtermBHs} shows the mass contribution of BHs and NSs to the bound mass as a function of time. The figure styles are as in Fig.~\ref{fig:mass}. In the model without kicks, more than 2 percent of cluster mass resides in BHs and more than one percent in NSs. Due to the absence of kicks, these fractions just reflect the birth rate of BHs and NSs, which are shown in green lines. If we use the updated SSE with fallback-scaled natal kicks, about one percent of the mass will remain in BHs, while NSs contribute ten times less. Any star that has a close encounter with a BH gains kinetic energy, and the BH decelerates, accelerating mass segregation. This manifests as a rapid increase of BH mass fraction shown in the top panel of Fig.~\ref{fig:longtermBHs}. At the end of the evolution of clusters with "old SSE, no kick", we can observe a rapid decrease of NS mass contribution in the bottom panel of Fig.~\ref{fig:longtermBHs}. This is a result of the ejection of NSs by BHs, as the BH subsystem forms in the center of the cluster.

\begin{figure}
	\includegraphics[width=\linewidth]{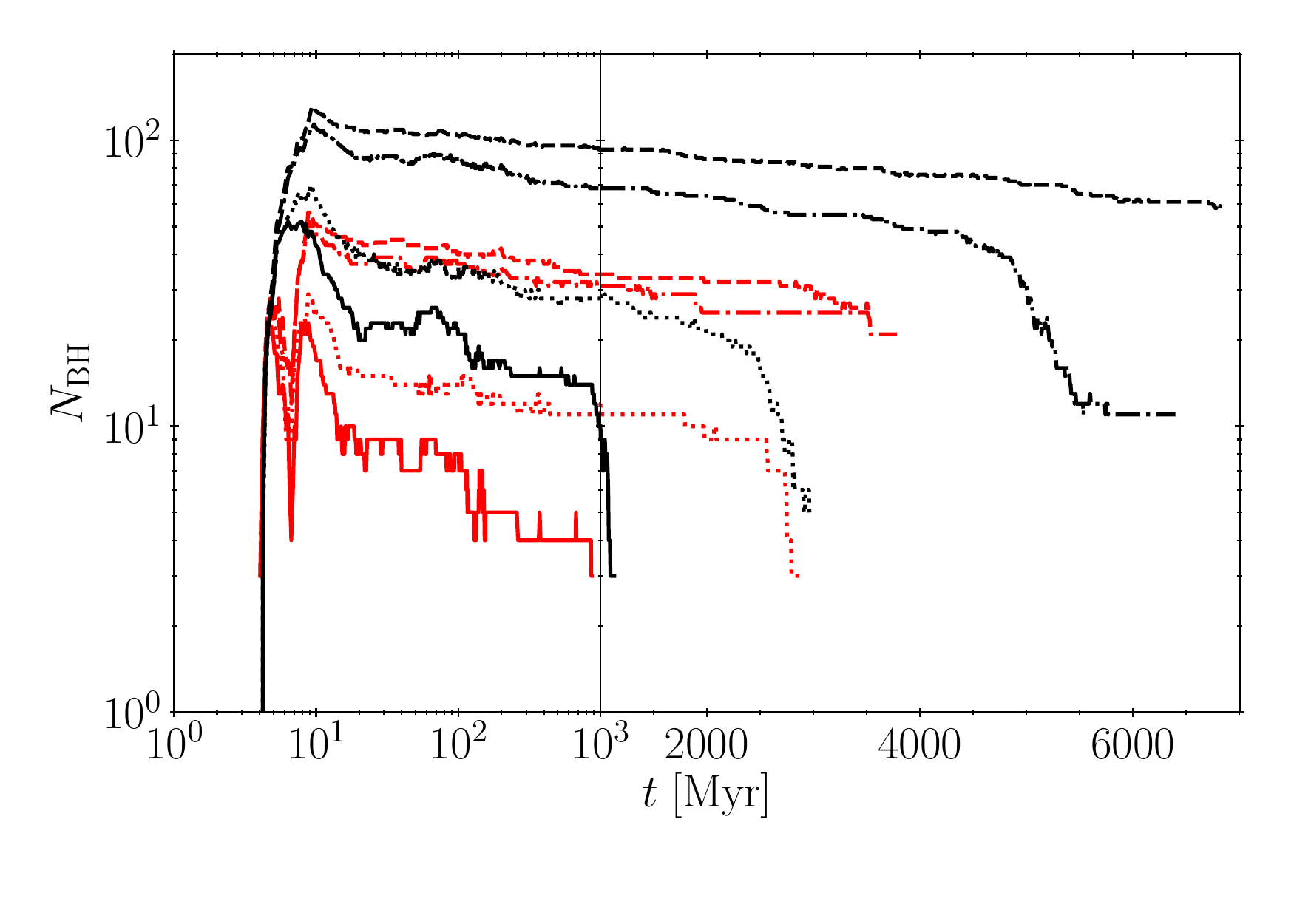}
    \caption{The evolution of the number of retained BHs for models with old SSE neglecting natal kicks and new SSE with fallback-scaled natal kicks. Colors correspond to the model sets, and line styles correspond to models with different $\sfe$ as Fig~\ref{fig:longtermBHs}. }
    \label{fig:nBH}
\end{figure}
\begin{figure}
	\includegraphics[width=\linewidth]{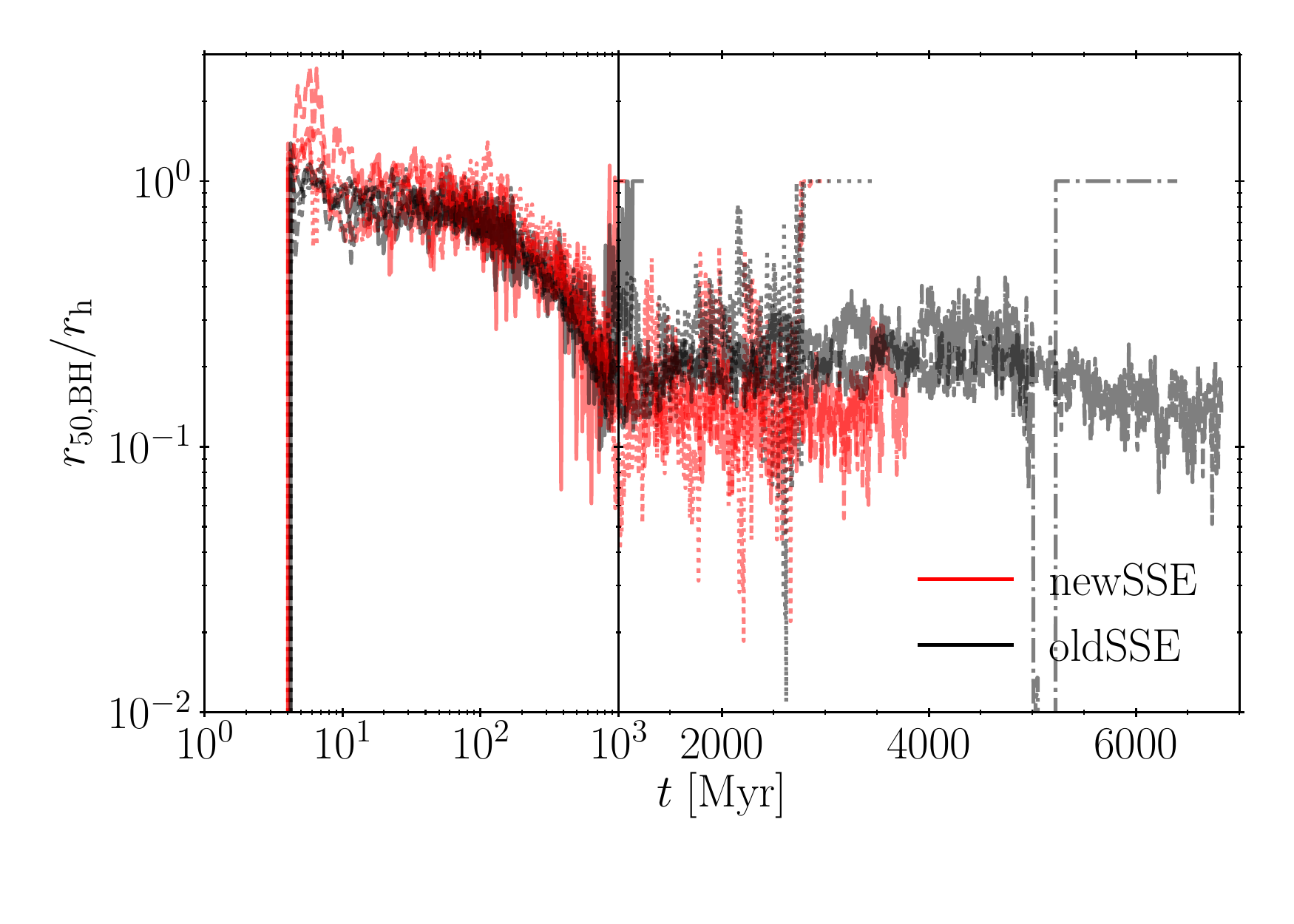}
    \caption{The evolution of half-mass radii of BH-sub-systems as fractions cluster half-mass radii for `old SSE, no kick' and `new SSE' models. Colors correspond to the model sets as indicated in the key. The line styles correspond to models with different $\sfe$. }
    \label{fig:r50}
\end{figure}

Figure~\ref{fig:nBH} shows the numbers of BHs residing within the Jacobi radius evolving through time. As we can see when we use the new SSE, the number of BHs in the cluster can reach $\sim 30$ depending on the value of $\sfe$. Fig.~\ref{fig:r50} shows the ratio of the half-mass radius for BHs to that of the bound cluster as a function of time. As we can see, this ratio is similar among all models retaining BHs. It is close to $1$ during the early evolution, meaning the BHs fraction is uniform across the cluster. The ratio then shrinks rapidly from $10^2$ to $10^3$ Myr after IGE, which is caused by mass segregation. At the end of the life only BHs are left in our model clusters indicated by the plateaus of $r_\mathrm{50,BH}/r_\mathrm{h}=1$. This is reminiscent of approaching a `dark cluster' phase \citep{Banerjee2011}.

{Such a formation of black hole sub-systems led to the creation of BH binary and multiple systems with small separations. Due to the softening of the gravitational potential, our simulations do not resolve the dynamics in the scale of the introduced softening length $\epsilon\approx41$~AU. In some simulations, BHs that accumulated to the cluster center became gravitationally bound despite the potential softening and could not escape each other. In that way, very tight BH systems formed in model clusters, which retained more than one BH.

The `old SSE, no kick' models have the most populated central BH systems. For example, the `old SSE, no kick' model cluster with $\sfe=0.25$ has a compact bound system of 16 BHs with a total mass of $150\msol$ in the cluster center at the age of 6.8 Gyr. 59 BHs and 191 NSs inhabit the $7000\msol$ cluster at this last available snapshot. The model cluster with $\sfe=0.20$ ends up as a system of 11 BHs with a total mass of $102\msol$. 
The model clusters with $\sfe=0.17$ and $\sfe=0.15$ have left $39.2\msol$ BH quartet and $17.2\msol$ BH binary, respectively. 

In the case of `new SSE,' only the model cluster with $\sfe=0.15$ has no BH binaries. The $\sfe=0.17$ model cluster has left a hierarchical triple system of BHs with a total mass of $28.3\msol$. The $\sfe=0.20$ cluster on its last available snapshot at 3.88 Gyr has two compact BBHs: One close to the cluster center, the other at almost 200 pc from the cluster in the Galactic field. At this time, the $M_\mathrm{J}=5480\msol$ cluster contains 21 BHs and 3 NSs. 


By compact BH systems, we mean the binary or multiple systems with a size comparable to the softening length scale, where we cannot resolve their dynamics.
Due to the potential softening, the BHs cannot experience close encounters to get enough acceleration for a recoil. Therefore, those compact BH systems cannot dissolve in our simulations. Instead, they oscillate in the harmonic potential of each other for a long time. 
}

In summary, our simulations show that even open clusters can form tight systems of BHs. Previous works achieved such tight systems only by assuming the presence of primordial binaries \citep{Banerjee2018,DiCarlo+2019} or by starting from dense systems in virial equilibrium \citep{Portegies_2000,Kumamoto+2019}. However, due to the softening of potential at small distances that we employ in our simulations, we cannot model the detailed evolution of the binary or multiple systems with our code. This will be the subject of future work.

\section{Conclusions}
\label{sec:con}

We studied the effect of different treatments of stellar evolution and supernova explosions (SSE) on the dynamics of star clusters with solar metallicity. We perform three sets of $N$-body simulations following an instantaneous gas expulsion. In the first set, we use the recent treatment of SSE that yields moderate kicks for supernova remnants due to the fallback of the ejecta. In the second set, we use an older version of SSE that results in strong kicks. Finally, we use the latter SSE with artificially removed kicks in the third set. We use star clusters formed with a constant star formation efficient (SFE) per free-fall time with a Plummer density profile at the time of the gas expulsion. Our model clusters move on a circular orbit in the Galactic disk plane at the solar distance (cf. Section \ref{sec:methods} for details).

The star clusters with smaller kick velocities retain a higher fraction of BHs, in agreement with previous studies \citep[e.g.][]{Kremer+2018,Pavlik+2018,Rodriguez+19,DiCarlo+2019,Kremer+2020ApJS..247...48K}(citations). In the model with strong kicks, no BHs remain in the stellar cluster. In contrast, our results show that it is possible to retain BHs in the cluster with the updated SSE without artificial reduction of the BH natal kick velocities. The contribution of BHs to the overall cluster mass is about $1\%$. This leads to forming a BH sub-system in the cluster center with a half-mass radius of $\sim 20\%$ of the bound cluster.    

Despite these differences, we find that  during the violent relaxation and the first Gyr of the evolution, the total bound mass of the clusters from different sets does not differ by more than $5\%$ with respect to each other. The difference mostly stems from the different fractions of retained BHs (and NSs) in clusters. Due to the similarity in the total mass, the overall dynamics of the clusters remain particularly close to each other in this early phase of the evolution.

However, the massive BHs in the stellar cluster lead to the continuous ejection of less massive objects from the cluster \citep{1993Natur.364..421K,1993Natur.364..423S}. In the longer term, this leads to dramatic differences between models with or without BHs. At the same time, the models that retain BHs (either by fallback-scaling or via artificial removal of the kicks) show similar dynamics even over longer time scales. For example, the artificial removal of kicks increases the retained BH mass fraction by only a few percent. 

To check the robustness of our results, we repeated our simulations for four different values of the global SFE. We find that our overall conclusions remain qualitatively the same for all SFE values.

In this work, we have not considered the effect of stellar binaries. The presence of primordial binaries does
influence the BH retention in clusters, as recent studies have shown \citep[e.g.,][]{Banerjee2018}, which is
worth exploring in a future study.

\section*{Acknowledgements}

This research has been funded by the Science Committee of the Ministry of Science and Higher Education of the Republic of Kazakhstan (Grant No. AP13067834) and the Nazarbayev University Faculty Development Competitive Research Grant Program (No 11022021FD2912). BS acknowledges funding the Aerospace Committee of the Ministry of Digital Development, Innovations and Aerospace Industry of the Republic of Kazakhstan (Grant No. BR11265408). The work of PB was supported by the Volkswagen Foundation under 
the special stipend No.~9B870 and the grant No.~97778.
PB acknowledges the support within the grant No.~AP14870501
of the Science Committee of the Ministry of Science and 
Higher Education of Kazakhstan. 
The work of PB was also supported under the special program of the NRF of Ukraine Leading and Young 
Scientists Research Support - ”Astrophysical Relativistic Galactic Objects (ARGO): life cycle 
of active nucleus”, No.~2020.02/0346.
SB acknowledges funding for this work
by the Deutsche Forschungsgemeinschaft (DFG, German Research Foundation)
through the project ``The dynamics of stellar-mass black holes in dense stellar systems and their
role in gravitational-wave generation'' (project number 405620641; PI: S. Banerjee).
BS acknowledges Dr. Taras Panamarev for his support and assistance in transferring the updated SSE into the phi-GRAPE/GPU code.

All simulations and data analysis have been carried out exclusively on high-performance workstations of ECL/NU\footnote{\href{https://sites.google.com/nu.edu.kz/ecl/facilities/computer-facilities/}{https://sites.google.com/nu.edu.kz/ecl/facilities/computer-facilities/}}.

\section*{Data Availability}

The data used in this work is available from the first author upon request.


\bibliographystyle{mnras}
\bibliography{myrefs} 








\bsp	
\label{lastpage}
\end{document}